# BRAIN COMPUTER INTERFACES FOR MOBILE APPS: STATE-OF-THE-ART & FUTURE DIRECTIONS

Sumit Soman, Siddharth Srivastava
*Department of Electrical Engineering*
*Indian Institute of Technology*
*Delhi, India*
*{eez127509, eez127506}@ee.iitd.ac.in*

Saurabh Srivastava
*Xerox Research Centre*
*Bangalore, India*
*saurabh.srivastava@xerox.com*

Nitendra Rajput
*IBM India Research Laboratory*
*New Delhi, India*
*rnitendra@in.ibm.com*

**ABSTRACT**

In recent times, there have been significant advancements in utilizing the sensing capabilities of mobile devices for developing applications. The primary objective has been to enhance the way a user interacts with the application by making it effortless and convenient. This paper explores the capabilities of using Brain Computer Interfaces (BCI), an evolving subset of Human Computer Interaction (HCI) paradigms, to control mobile devices. We present a comprehensive survey of the state-of-the-art in this area, discussing the challenges and limitations in using BCI for mobile applications. Further we propose possible modalities that in future can benefit with BCI applications. This paper consolidates research directions being pursued in this domain, and draws conclusions on feasibility and benefits of using BCI systems effectively augmented to the mobile application development domain.

## 1. INTRODUCTION

Advancements in mobile hardware and sensing capabilities have drastically altered the scope of using mobile devices as well as the procedure in which the user interacts with them. Multiple communication modalities, effective HCI paradigms and a versatile communication network have paved the way for building mobile applications which interact with the user in a more "natural" manner. The next development appears to be the use of consciously modulated parameters of the human body to communicate with mobile apps, which would make user interaction seamless and intuitive. As one such use case, BCI as a means for HCI on mobile devices is beneficial to enable communication for patients with locomotor disabilities. BCI presents a viable alternative, as conventional communication channels are not available for interacting with mobile platforms. Drawing motivation from such scenarios, we seek to explore the state-of-the-art in the use of BCI as a modality for providing user input to mobile applications.

There are several challenges in building mobile applications based on BCI systems. A key challenge is the choice of control paradigm, or precisely, what actions should the user perform which would be detected by the BCI system to execute actions on the mobile application. This is a difficult choice due to the variations in reproducibility of the control intents across users, which directly affects the classification accuracy of the BCI system. Another important challenge is the platform for processing of EEG signals. The EEG signals from acquisition devices are represented as floating point matrices, which have high dimensionality. To provide a representative estimate, EEG data acquired from 32 channels (recording locations) for 5 seconds at a sampling rate of 250 Hz would be represented by a matrix of size [$32 \times 1250$]. Operations on large matrices require high computational cost. However, the computational power available on mobile platforms is limited, hence the signal-processing pipeline needs to be optimized to suit the platform architecture. The challenges in the use of BCI for gaming applications have been discussed in the work by Nijholt et al. [40, 39, 38]. Despite these, BCIs have also been used in developing assistive technology [8], mobile robot control [3, 37] and many other applications [36, 23, 19].

The objective of this paper is to present a survey of the recent work done in the field of applications developed for mobile platforms based on BCI. To the best of our knowledge, this would be a pioneering effort in this direction, which will enable consolidation of research being pursued and draw conclusions on feasible future directions. The rest of the paper is organized as follows. Section 2 discusses the details of the survey methodology used in this paper, followed by the characteristics required in an EEG acquisition system feasible for development of mobile applications in Section 3. This is followed by a discussion on the modalities employed in developing BCI based mobile applications along with a few BCI based mobile applications in Section 4. The conclusions and future directions are presented in Section 5.

## 2. SURVEY METHODOLOGY

The content presented in this paper is based on research papers and articles that are indexed and available on Google Scholar and Scopus. Papers with index terms using combinations of "Mobile" with "BCI", "Assistive Technology", "EEG", "Brain Machine Interface (BMI)" and "HCI" have been discussed in the review. The focus has been to lay more emphasis on papers, which have appeared in the past 3 years, though this has not been strictly adhered to in cases where seminal work has been indicated.

## 3. BUILDING BCI SYSTEMS FOR MOBILE: DEVICES & MODALITIES

Developing BCI systems for practical mobile applications involves several challenges owing to portability and degree of involvement of the user in triggering control intents. Conventional EEG acquisition devices need to be made portable and ergonomic, while control paradigms should need minimal effort and not cause fatigue. This section presents details on characterizing an "ideal" acquisition device, and discusses some of the common paradigms used in such applications.

### 3.1 EEG Acquisition Systems for Mobile

One of the primary challenges faced while using BCI for mobile platforms has been the choice of the EEG Acquisition System. Conventional EEG acquisition systems suffer with the following constraints from the perspective of usability and flexibility over mobile devices:
1. They are large and non-portable.
2. They have large number of electrodes (EEG acquisition sensors).
3. Many interconnecting wires to connect the amplifier with the electrodes.
4. Requirement of a trained professional for connecting electrodes as per the 10-20 system [22].
5. Usage requires application of a conductive gel [50] or saline.

Such systems are more suited for EEG-based investigations in medical facilities. Due to these constraints, they cannot be directly used for mobile-based EEG applications. However, the benefit lies in the higher resolution of the EEG signals obtained from them in terms of number of channels, lower noise levels, higher sampling rates and the ability to configure the channel montage. This in turn aids the machine learning pipeline to achieve higher classification accuracies. Hence, an ideal mobile EEG acquisitions device would have a compact form factor, be easy to wear and setup, lightweight and enable real-time processing of EEG signals. It would have a convenient and wireless connection interface, preferably over Bluetooth or Wi-Fi. High-quality signals should be obtained, in terms of frequency resolution and low-noise.

In view of the challenges identified above, newer EEG systems have been developed. Gargiulo et al. [14] demonstrate an EEG system, called Penso, with a compact form factor and Bluetooth connectivity. The system has shown high correlation coefficient of 0.83 on average with clinical grade EEG in terms of alpha [27] and mu [42] rhythm detection for 8 subjects. It has dry electrodes made with conducting rubber, which enable it to be setup easily. The device has a measurement bandwidth of 0.4-40 Hz, and could be setup with an average setup time of 10 seconds per electrode. This is lower than the time required for a system with wet electrodes, which would be around 2-3 minutes per electrode. This reduction in setup time is significant.

Further, Chi et al. [7] have developed a dry and contact-less EEG system for use in mobile applications. These electrodes do not require direct contact with the scalp and are based on capacitive coupling with the scalp through the hair. They also have active buffering capability; high impedance and effective shielding against noise, which provides improved quality EEG signals. The developed system has been shown to achieve an Information Transfer Rate (ITR) of 19 bits/minute with 100% accuracy for the SSVEP modality. Another popular mobile EEG headset has been the Emotiv EPOC headset [48]. It is a 14 channel wireless headset with saline-based electrodes. The headset has been ergonomically designed and is easy to wear and setup as it has a wireless interface. It has been used in a variety of gaming applications [53, 44] and assistive technologies [30, 17]. The headset also has a more compact version called the "Emotiv Insight" [1]. Other portable systems include the Neurosky and G.MOBIlab [16]. A review of the transition from laboratory to mobile BCI has been discussed in the work by Edlinger [13], and requirements of a portable system have been enumerated by Debener et al. [12]. A review of EEG systems used for BCI can be found in the work by Lin et al. [32].

## 3.2 BCI Control Modalities for Mobile Applications

This section aims to discuss the modalities available and in use for developing mobile applications based on BCI. We consider gestures and eye movements, SSVEPs and P300 as primary control paradigms.

### 3.2.1 Gestures and Eye Movements

The ability to detect facial gestures, eye-blinks and eye movements from the EEG signals makes it a convenient paradigm to enable control in mobile applications. This modality encompasses a variety of control triggers, including eye blinks, lateral eye movements, facial gestures such as wink, twitch, jaw clench and smirk. The benefit of using these lies in the fact that a variety of intents are available, each of which translates into direct actions. Hence, it offers higher degree of freedom in terms of multi-class BCI. These can also be integrated with other modalities like eye gaze tracking [49].

### 3.2.2 SSEVP

Steady State Visually Evoked Potentials (SSVEPs) have been used to develop several BCI systems. The key concept is that the user's brain responds to stimulus presented at specific frequencies, which may be realized by flashing the commands to the user at different frequencies. When the user focuses attention on the object of interest, the brain will generate signals at that frequency. The advantage is that this is less susceptible to artifacts, and reproducible across subjects [2, 54]. The downside is that users may find the stimulus overwhelming, as it needs to be flashed at different frequencies. For instance, [60] indicates that the use of SSVEPs may induce seizures or impair vision. Recent works on adapting SSVEPs include the development of stimulus presentation paradigms for mobile applications [56] and testing the versatility of SSVEPs while users are engaged in other physical activities [34]. In [35], the authors present a feasibility study of using SSVEPs when the subject is engaged in walking. This shows promise in the use of this paradigm. Examples of SSVEP based BCI applications for mobile have been discussed in Section 5.

### 3.2.3 P300

The P300 signal is another well-known BCI control intent. It represents EEG signals from the parietal lobe, which is known to occur 300 milliseconds after an "odd-ball" event. This has been used in the P300 speller, where the user is shown a grid of characters in which the rows and columns are flashed in a predetermined sequence. The user is asked to focus on the letter of interest and count the number of times it is flashed. This is termed as an "odd-ball" event, which leads to the generation of a P300 signal. It can be detected by EEG signal processing to identify the user intent. In this manner, users are able to spell words using the P300 speller. Efforts have been made to study the versatility of using P300 [43], in the context of mobile applications as an extension of the P300 speller with predictive text [21], and in combination with auditory stimulus by Vos et al. [10, 11] Their study for a 3-class auditory oddball task on 20 subjects gave average classification accuracies of 71% for sedentary and 64% for walking states. The task used a standard tone of 900 Hz and two deviant tones of 600 and 1200 Hz with a mean inter-stimulus interval of 1 second and jitter between 0-375ms. The data analysis involved re-referencing, Extended Infomax ICA [28], and classification using Stepwise LDA [25].

Notable conclusions and inferences were presented by the authors in [10,11]. They opined that practical BCI systems should be robust, reliable and easy to setup outside the controlled laboratory environment. It has been inferred that lower accuracies of BCI systems while subjects are performing physical actions can be attributed to the fact that the brain is involved with other tasks. As P300 amplitudes are a measure of the attention levels, these decrease in such cases. An important proposition to resolve this is multi-modal BCI, which incorporates feedback from other paradigms such as eye tracking, head movement (gyroscope) and video recording of visual scene. However, the practical realization of any multi-modal system would also demand computationally intensive processing. Hence, there is a need to develop such a processing pipeline, which is an active area of research. For instance, minimal complexity classifiers have been investigated in [26, 24].

## 4. BCI FOR MOBILE APPS: STATE OF THE ART

The SmartPhone Brain Scanner (SBS) app developed by Stopczynski et al. [45, 46, 47] provides a hand-held brain scanner on mobile using the Emotiv headset for real-time image reconstruction of brain activity. This application serves as an assistive tool for EEG diagnosis in remote areas. The open source implementation allows extension of the framework to other BCI-based mobile applications. It provides multiple feature extraction and classification methods, making this a candidate platform for BCI-based mobile application development. As the authors indicate in [45], SBS does not have automatic artifact detection or elimination methods, which are crucial for efficient EEG signal processing. On similar lines, Lin et al. [33] perform remote transmission of brain images over communication networks using client-server architecture. The objective is a mobile visualization tool, with focus on efficient structures for storage, transmission and re-construction of the brain images over the network to remote devices.

We now discuss day-to-day applications based on BCI for mobile platforms. Neurophone, by Campbell et al. [6] is a phone-dialer application based on the P300 paradigm. Users are shown images of contacts flashed randomly and the P300 stimulus is used as a trigger to initiate a call with the contact. Their paper presents two versions of this application, one that is triggered by eye-winks and another that is triggered using P300. The paper concludes that the "wink-triggered" version works robustly even in noisy conditions, whereas the "thought-triggered" dialer appears "promising but at present less reliable." The application has been tested on three subjects under sedentary and walking conditions, and reports accuracies in the range of 92-95% for the "wink-triggered" version and lower accuracies for the P300 version. This work is novel as a working application has been developed and evaluated rigorously.

A P300 based BCI Messenger has been developed by Li et al. [29] as a tool for day-to-day communication, and has achieved over 80% accuracy on 6 subjects across sessions. The application consisted a $[6 \times 6]$ grid with Chinese strokes to be used as a virtual keyboard. The training phase involved the subjects viewing specific targets on the screen while their EEG was being recorded, and a subsequent phase where visual feedback was provided to this task. The subjects undertook 3 sessions of training with around 20 runs per session. The paper reports an average accuracy of 69.7%, and suggests possible improvement by use of larger-sized visual stimulus, use of channel selection to reduce EEG dimensionality and use of alternate classifiers. The *CharStreamer* paradigm proposed by Höhne et al. [20] aims to incorporate auditory and visual stimulus in order to realize a speller using Event Related Potentials (ERPs).

Several BCI based mobile applications using the SSVEP paradigm are available in literature. Wei et al. [58] develop an SSVEP based user calling application with an LED-based visual simulator panel for eliciting user input. The input is mapped to key-presses on the standard numeric keypad with dedicated keys for call receiving/disconnecting. The user interface consists of a [4 × 4] grid with 10 number keys and 6 function keys for dial, disconnect, answer, fast dial, enter (to complete current operation) and backspace.

Table 1. A comparison of recent works in the area of BCI-based applications for mobile platforms

| S.No. | Application | BCI Paradigm | EEG Headset | Mobile Device | Preprocessing &Feature | Classification | Remarks |
|---|---|---|---|---|---|---|---|
| 1 | Neurophone [6] | P300 | Emotiv | I-Phone | Band Pass Filtering | Multivariate Equi-prior | Contact caller App |
| 2 | Smartphone Brain Scanner | - | Emotiv | Multi-platform | Multiple - Fast Fourier | Bayesian Formulations, | Brain activity Imagery App |
| 3 | Cell phone BCI [55] | SSVEP | MEMS Sensors | Nokia/ J2ME | FFT, Canonical | Information Transfer Rate | Phone Dialer |
| 4 | BCI Messenger [29] | P300 | Neuroscan QuickCap | - | Artifact Removal by | Template Matching, | Chinese and English |
| 5 | Char Streamer [20] | ERP | EasyCap GmbH | - | Artifact Removal and | Binary Linear Discriminant | Spelling interface |
| 6 | BCI Telepresence | Motor Imagery | G.Tec | Robotino | Laplacian spatial filtering, | Statistical Gaussian | Telepresence mobile robot |
| 7 | Hex-o-spell [4] | Motor Imagery | Berlin BCI | - | CSP (as in BBCI) | LDA (as in BBCI) | Brain actuated |
| 8 | Mental Telephone | SSVEP | Self made | SIM300 | CCA | Based on CCA coefficients | Phone dialer app |
| 9 | Unlock Project [5] | SSVEP | G.Tec, Emotiv, | Android | Not Specified | Not Specified | Framework for |
| 10 | Mobile Device Control [52] | S-EMG | Motion Lab Systems | Android | Band Pass Filtering& | Weighted bandpower | Mobile app to control |
| 11 | Mobile Controller [9] | S-EMG | Arm-band for EMG | | Not specified | Not specified | Subtle Gesture based device |

The system is based on detecting SSVEPs from occipital electrodes, and uses frequency screening to identify feasible frequency bands for specific users. During training, users are presented stimuli at frequencies of 6-20 Hz with a window of 0.5 Hz, and an optimal band is identified using canonical correlation analysis. The developed system has been tested on nine volunteers and reports mean ITR of 33.53 bits/min, which can be attributed to the training process that users undergo prior to online system usage.

Wang et al. [56] design a Phone Dialer app with an average ITR of 26.46 bits/min. A general platform for developing SSVEP based BCI systems with minimal delay has also been developed by Brumberg et al. [5]. It supports multiple EEG acquisition devices and provides a platform agnostic framework. It is evident from the work in this direction that SSVEP based BCI mobile applications have high potential in the future. Ideas for shared brain control in mobile platforms has been explored in the work by Tonin et al. [51]. BCI can also be integrated in a multimodal fashion for interaction with mobile devices [18]. More recently, Lin et al. [31] have demonstrated the use of Micro-Electro Mechanical Systems (MEMS) based EEG sensors to realize wireless and non-invasive prosthesis control. A review of application of BCIs for control of mobile peripherals has been presented by Li et al. [41]. Work has also been done using modalities other than EEG signals, such as the Surface Electromyogram (S-EMG), for gesture based device control [9] and household electronic devices control [52]. A summary is given in Table 1.

## 5. DISCUSSION & FUTURE DIRECTIONS

This paper presented a survey of BCI based applications developed for mobile platforms. Despite several challenges, BCI research has evolved to many practical applications. This modality offers a natural extension to how humans interact with the external environment, and hence presents a promising avenue.

Figure 1 shows a graph of the number of research articles on mobile-based BCI over the last ten years. It is evident that the trend indicates increasing favor of these research directions, especially over the last six years. Other avenues of interest appear to be developing new paradigms for the mobile devices using EEG signals, for instance the P300 speller [15]. Extensions such as the mental typewriter Hexospell [4] can be developed for mobile platforms, based on probabilistic [59] or multi-modal [57] text entry. SSVEP based applications hold a promising future, owing to the natural adaptability of this paradigm to a menu-based interface. For example, the home menu on any mobile platform with options for contacts, messaging, games, settings, gallery and the like can be implemented as a grid of flashing options, where selection is triggered by SSVEP.

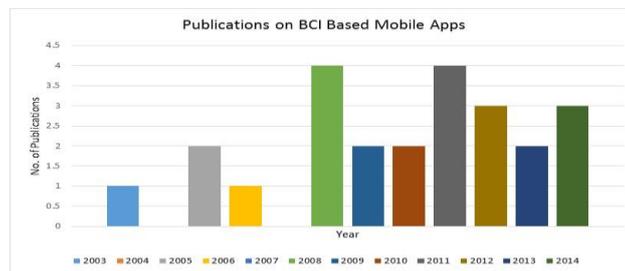

Figure 1. Research papers published over the last ten years in the area of developing mobile applications using BCI.

Mobile devices have several sensors such as accelerometer, gyroscope, compass, barometer, proximity sensors etc., which provide data that can be operated upon by machine learning techniques to augment the user intent. This includes techniques for processing images from camera feed or audio processing from the microphone. The combination of these can be used as an effective measure against the false-positives identified by the BCI modality. User-specific data is also available on mobile devices, which can be harvested to augment BCI-based techniques. Further, data from of mobile usage patterns such as touch, call, applications etc. can be used based on the problem at hand. For instance, the most likely contacts whom the user may call can be determined using call logs and the user can be presented these callers to choose from using the BCI modality (as the first level of choice). The use of BCI for mobile holds promise, as it would be useful for developing both assistive and general entertainment technology for the evolving mobile user. Developments in the fields of computational neurology, signal processing, machine learning coupled with the increasing availability of computational power on the mobile device along with cloud storage and processing paves the way for development of more BCI based mobile applications for the ordinary user.

## REFERENCES

[1] (2014) Emotiv insight neuroheadset URL - *http://www.emotiv.com/insight.php*

[2] Allison B., Luth T., Valbuena D., et al, (2010) BCI demographics: How many (and what kinds of) people can use an SSVEP BCI? *IEEE Transactions on Neural Systems and Rehabilitation Engineering,* 18(2):107–116.

[3] Bell CJ., Shenoy P., Chalodhorn R., et al, (2008) Control of a humanoid robot by a noninvasive brain–computer interface in humans .*Journal of neural engineering* 5(2):214.

[4] Blankertz B., Dornhege G., Krauledat M., et al, (2006) The Berlin brain-computer interface presents the novel mental typewriter hex-o-spell.

[5] Brumberg JS., Lorenz SD., Galbraith BV.et al, (2012) The unlock project: A python-based framework for practical brain-computer interface communication app development. *Engineering in Medicine and Biology Society (EMBC), 2012 Annual International Conference of the IEEE, IEEE*, pp 2505–2508.

[6] Campbell A, Choudhury T, Hu S, et al, (2010) Neurophone: brain-mobile phone interface using a wireless EEG headset. *Proceedings of the second ACM SIG- COMM workshop on Networking, systems, and applications on mobile handhelds*, ACM, pp 3–8.

[7] Chi Y, Wang YT, Wang Y, et al, (2012) Dry and noncontact EEG sensors for mobile brain computer interfaces. *IEEE Transactions on Neural Systems and Rehabilitation Engineering*, 20(2):228–235.

[8] Cincotti F, Mattia D, Aloise F, et al, (2008) Non-invasive brain–computer interface system: towards its application as assistive technology. *Brain research bulletin* 75(6):796–803.

[9] Costanza E, Inverso SA, Allen R (2005) Toward subtle intimate interfaces for mobile devices using an EMG controller. *Proceedings of the SIGCHI conference on Human factors in computing systems*, ACM, pp 481–489.

[10] De Vos M, Gandras K, Debener S (2014) Towards a truly mobile auditory brain–computer interface: Exploring the p300 to take away. *International Journal of Psychophysiology* 91(1):46–53.

[11] De Vos M, Kroesen M, Emkes R, et al, (2014) P300 speller BCI with a mobile EEG system: comparison to a traditional amplifier. *Journal of neural engineering* 11(3):2014.

[12] Debener S, Minow F, Emkes R, Gandras K, Vos M (2012) How about taking a low-cost, small, and wireless EEG for a walk? *Psychophysiology* 49(11):1617–1621.

[13] Edlinger G, Krausz G, Laundl F, et al, (2005) Architectures of laboratory-pc and mobile pocket pc brain-computer interfaces. *Proceedings of 2nd International IEEE EMBS Conference on Neural Engineering, 2005*., pp 120–123.

[14] Gargiulo G, Bifulco P, Calvo R, Cesarelli M, Jin C, van Schaik A (2008) A mobile EEG system with dry electrodes. *Biomedical Circuits and Systems Conference (BioCAS) 2008.* IEEE, pp 273–276.

[15] Guan C, Thulasidas M, Wu J (2004) High performance p300 speller for brain-computer interface. *IEEE International Workshop on Biomedical Circuits and Systems, 2004, IEEE*, pp S3–5.

[16] Guger C., et al, P300 spelling device with g. mobilab and Simulink.

[17] Gupta S, et al, (2012) Detecting eye movements in EEG for controlling devices. *IEEE International Conference on Computational Intelligence and Cybernetics (CyberneticsCom)*, 2012, pp 69–73.

[18] Gürkök H, Nijholt A (2012) Brain–computer interfaces for multimodal interaction: a survey and principles. *International Journal of Human-Computer Interaction* 28(5):292–307.

[19] Hintermüller C, Kapeller C, Edlinger G, Guger C (2013) BCI integration: Application interfaces.

[20] Höhne J, Tangermann M (2014) Towards user-friendly spelling with an auditory brain-computer interface: The charstreamer paradigm. *PLOS ONE* 9(6):e98,322.

[21] Hohne J, Schreuder M, Blankertz B, Tangermann M (2010) Two-dimensional auditory P300 speller with predictive text system. *Annual International Conference of the IEEE Engineering in Medicine and Biology Society (EMBC)*, 2010, pp 4185–4188.

[22] Homan RW, Herman J, Purdy P (1987) Cerebral location of international 10–20 system electrode placement. *Electroencephalography and clinical neurophysiology* 66(4):376–382.

[23] Huggins JE, Guger C, Allison B, et al (2014) Brain- Computer Interfaces, *In the Workshops of the fifth international brain-computer interface meeting: Defining the future.* 1(1):27–49.

[24] Jayadeva. Learning a hyperplane classifier by minimizing an exact bound on the VC dimension." *Neurocomputing* 149 (2015): 683-689, DOI http://dx.doi.org/10.1016/j.neucom.2014.07.062

[25] Jennrich RI, Sampson P (1977) *Stepwise discriminant analysis. Statistical methods for digital computers* 3:77–95.

[26] Keerthi SS, Chapelle O, DeCoste D (2006) Building support vector machines with reduced classifier complexity. *The Journal of Machine Learning Research* 7:1493–1515.

[27] Klimesch W (1999) EEG alpha and theta oscillations reflect cognitive and memory performance: a review and analysis. *Brain research reviews* 29(2):169–195.

[28] Lee TW, Girolami M, Sejnowski TJ (1999) Independent component analysis using an extended infomax algorithm for mixed subgaussian and supergaussian sources. *Neural computation* 11(2):417–441.

[29] Li Y, Zhang J, Su Y, Chen W, Qi Y, Zhang J, et al, (2009) P300 based BCI messenger. *In proceedings of ICME International Conference on Complex Medical Engineering, 2009*. pp 1–5.

[30] Lievesley R, Wozencroft M, Ewins D (2011) The Emotiv EPOC neuroheadset: an inexpensive method of controlling assistive technologies using facial expressions and thoughts. *Journal of Assistive Technologies* 5(2):67–82.

[31] Lin CT, Ko LW, et al, (2008) Noninvasive neural prostheses using mobile and wireless EEG. *Proceedings of the IEEE* 96(7):1167–1183.

[32] Lin CT, Ko LW, Chang MH, et al, (2009) Review of wireless and wearable electroencephalogram systems and brain-computer interfaces–a mini-review. *Gerontology* 56(1):112–119.

[33] Lin MK, Nicolini O, Waxenegger H, Galloway GJ, Ullmann JF, Janke AL (2013) Interpretation of medical imaging data with a mobile application: a mobile digital imaging processing environment. *Frontiers in neurology*.

[34] Lin YP, Wang Y, Jung TP (2013) A mobile SSVEP based brain-computer interface for freely moving humans: The robustness of canonical correlation analysis to motion artifacts. *35th Annual International Conference of the IEEE Engineering in Medicine and Biology Society (EMBC),* 2013, IEEE, pp 1350–1353.

[35] Lin YP, Wang Y, Jung TP (2014) Assessing the feasibility of online SSVEP decoding in human walking using a consumer EEG headset. *Journal of neuro-engineering and rehabilitation* 11(1):119.

[36] Millán JdR (2013) Brain-computer interfaces. *Introduction to Neural Engineering for Motor Rehabilitation* 40:237.

[37] Millán JdR, Renkens F, Mourino J, et al, Non-invasive brain-actuated control of a mobile robot. *Proceedings of the 18th international joint conference on Artificial intelligence*, (2003) pp 1121–1126.

[38] Nijholt A (2009) BCI for games: A state of the art survey. *Entertainment Computing-ICEC 2008*, Springer, pp 225–228.

[39] Nijholt A, van Erp JB, Heylen D (2008) *Braingain: BCI for HCI and games.*

[40] Nijholt A, Bos DPO, Reuderink B (2009) Turning shortcomings into challenges: Brain–computer interfaces for games. *Entertainment Computing* 1(2):85–94.

[41] Penghai L, DongDHWBM (2011) Research progress on application of brain-computer-interface in mobile peripheral control. *Journal of Biomedical Engineering* 3:038.

[42] Pfurtscheller G, Brunner C, Schlögl A, Lopes da Silva F (2006) Mu rhythm (de) synchronization and EEG single-trial classification of different motor imagery tasks. *Neuroimage,* 31(1):153–159.

[43] Polich J (1997) On the relationship between EEG and P300: individual differences, aging, and ultradian rhythms. *International journal of psychophysiology* 26(1-3):299–317.

[44] Scherer R, Proll M, Allison B, Muller-Putz GR (2012) New input modalities for modern game design and virtual embodiment. *Virtual Reality Short Papers and Posters (VRW),* 2012 IEEE, IEEE, pp 163–164.

[45] Stopczynski A, Larsen JE, Stahlhut C, et al, (2011) A smartphone interface for a wireless EEG headset with real-time 3D reconstruction. *Affective Computing and Intelligent Interaction*, Springer, pp 317–318.

[46] Stopczynski A, et al, (2014) The smartphone brain scanner: A portable real-time neuroimaging system. *PloS one* 9(2):e86,733.

[47] Stopczynski A, et al, (2014) Smartphones as pocketable labs: Visions for mobile brain imaging and neurofeedback. *International Journal of Psychophysiology* 91(1):54–66.

[48] Stytsenko K, Jablonskis E, Prahm C (2011) Evaluation of consumer EEG device Emotiv EPOC. *Proceedings of MEi: CogSci Conference 2011*, Ljubljana.

[49] Takahashi K, Ohta T, Hashimoto M (2008) Remarks on EOG and EMG gesture recognition in hands-free manipulation system. In: Systems, Man and Cybernetics, 2008. SMC 2008., pp 798–803

[50] Tallgren P, Vanhatalo S, Kaila K, Voipio J (2005) Evaluation of commercially available electrodes and gels for recording of slow EEG potentials. *Clinical Neurophysiology* 116(4):799–806.

[51] Tonin L, Leeb R, Tavella M, Perdikis S, del Millan J (2010) The role of shared-control in BCI-based telepresence. *In proceedings of IEEE International Conference on Systems Man and Cybernetics* (SMC), 2010, pp 1462–1466.

[52] Vernon S, Joshi SS (2011) Brain–muscle–computer interface: mobile-phone prototype development and testing. *IEEE Transactions on Information Technology in Biomedicine*, 15(4):531–538.

[53] Van Vliet M, et al, (2012) Designing a brain-computer interface controlled video-game using consumer grade EEG hardware. *Proceedings of Biosignals and Biorobotics Conference (BRC),* 2012 ISSNIP, IEEE, pp 1–6.

[54] Volosyak I, et al, (2011) BCI demographics II: how many (and what kinds of) people can use a high-frequency SSVEP BCI? *IEEE Transactions on Neural Systems and Re- habilitation Engineering*, 19(3):232–239.

[55] Wang YT, Wang Y, Jung TP (2011) A cell-phone-based brain–computer interface for communication in daily life. Journal of neural engineering 8(2):025,018.

[56] Wang YT, Wang Y, Cheng CK, Jung TP (2013) Developing stimulus presentation on mobile devices for a truly portable SSVEP-based BCI. *Proceeding of 35th Annual International Conference of the IEEE Engineering in Medicine and Biology Society (EMBC),* 2013, IEEE, pp 5271–5274.

[57] Ward DJ, Blackwell AF, MacKay DJ (2000) Dashera data entry interface using continuous gestures and language models. *Proceedings of the 13th annual ACM symposium on User interface software and technology, ACM*, pp 129–137.

[58] Wei Q, Zou X, Lu Z, Wang Z (2014) Design and implementation of a mental telephone system based on steady-state visual evoked potential. *Journal of Computational Information Systems* 10(2):547–554.

[59] Williamson J, Murray-Smith R (2005) Hex: Dynamics and probabilistic text entry. *Switching and Learning in Feedback Systems, Springer,* pp 333–342.

[60] Zhu D, Bieger J, Molina GG, Aarts RM (2010) A survey of stimulation methods used in SSVEP-based BCIs. *Computational intelligence and neuroscience* 2010.